\documentclass[conference]{IEEEtran}
\IEEEoverridecommandlockouts
\usepackage{cite}
\usepackage{amsmath,amssymb,amsfonts}
\usepackage{algorithmic}
\usepackage{graphicx}
\usepackage{textcomp}
\usepackage{xcolor}
\newcommand{\framework}{VeriPPA\xspace}
\usepackage{url}
\usepackage{hyperref}

\makeatletter
\renewcommand\footnoterule{%
  \kern \p@ 
  \hrule width 0.65\linewidth height 0.4pt
  \kern-5\p@ 
}
\makeatother

\usepackage{array}
\usepackage{booktabs} 
\usepackage[boxed,ruled]{algorithm2e}
\usepackage{multirow}

\def\BibTeX{{\rm B\kern-.05em{\sc i\kern-.025em b}\kern-.08em
    T\kern-.1667em\lower.7ex\hbox{E}\kern-.125emX}}

\begin{document}

\title{LLM-VeriPPA: Power, Performance, and Area Optimization aware Verilog Code Generation with Large Language Models
{\footnotesize \textsuperscript{}}
\thanks{}
}


\author{%
Kiran Thorat\IEEEauthorrefmark{1}\textsuperscript{,+},
Jiahui Zhao\IEEEauthorrefmark{1}\textsuperscript{,+},
Yaotian Liu\IEEEauthorrefmark{2},
Amit Hasan\IEEEauthorrefmark{1},
Hongwu Peng\IEEEauthorrefmark{1},
Xi Xie\IEEEauthorrefmark{1},
Bin Lei\IEEEauthorrefmark{3}, 
Caiwen Ding\IEEEauthorrefmark{3}
\\[1ex]
\IEEEauthorblockA{%
\IEEEauthorrefmark{1}University of Connecticut,
\{kiran\_gautam.thorat, jiahui.zhao, amit.hasan,  hongwu.peng, xi.xie\}@uconn.edu}
\IEEEauthorblockA{%
\IEEEauthorrefmark{2}Arizona State University,
yaotian\_liu@asu.edu}
\IEEEauthorblockA{%
\IEEEauthorrefmark{3}University of Minnesota, 
\{bin.lei, dingc\}@umn.edu}
\thanks{\textsuperscript{+} These authors contributed equally}
}
\maketitle

\begin{abstract}
Large Language Models (LLMs) are gaining prominence in various fields, thanks to their ability to generate high-quality content from human instructions. This paper delves into the field of chip design using LLMs, specifically in Power-Performance-Area (PPA) optimization and the generation of accurate Verilog codes for circuit designs. We introduce a novel framework \framework designed to optimize PPA and generate Verilog code using LLMs. Our method includes a two-stage process where the first stage focuses on improving the functional and syntactic correctness of the generated Verilog codes, while the second stage focuses on optimizing the Verilog codes to meet PPA constraints of circuit designs, a crucial element of chip design. 
Our framework achieves an 81.37\% success rate in syntactic correctness and 62.06\% in functional correctness for code generation, outperforming current state-of-the-art (SOTA) methods. On the RTLLM dataset. On the VerilogEval dataset, our framework achieves 99.56\% syntactic correctness and 43.79\% functional correctness, also surpassing SOTA, which stands at 92.11\% for syntactic correctness and 33.57\% for functional correctness. Furthermore, Our framework able to optimize the PPA of the designs.
These results highlight the potential of LLMs in handling complex technical areas and indicate an encouraging development in the automation of chip design processes. Our source codes are here.
\footnote{\url{https://github.com/kiranthorat3/LLMVeriPPA}}
\IEEEpubidadjcol
\end{abstract}
\section{Introduction}



As Moore’s law continues to drive design complexity and scaling in chip design, it pushes chip design tools like Electronic Design Automation (EDA) to their limits. These traditional tools are time-consuming and rely on human experts.
Machine learning (ML) has been successfully integrated into chip design for logic synthesis \cite{8351885, 9045559}, placement \cite{10.1145/2228360.2228497}, routing \cite{10.1145/3372780.3375560, 8533535}, testing \cite{10.1145/2429384.2429404}, and verification \cite{10.1145/775832.775907, 10.1145/3195970.3196059}. 
The popularity of agile hardware design exploration has been on the rise due to the growth of large language models (LLMs). A promising direction is using natural language instruction to generate hardware description language (HDL), e.g., Verilog, aiming to greatly lower hardware design barriers and increase design productivity, especially for users who do not possess extensive expertise in chip design \cite{10720939, pei2024bettervcontrolledveriloggeneration, pinckney2024revisiting, Wu_2024}.

\begin{figure*}
    \centering
     \includegraphics[width=0.95\textwidth]{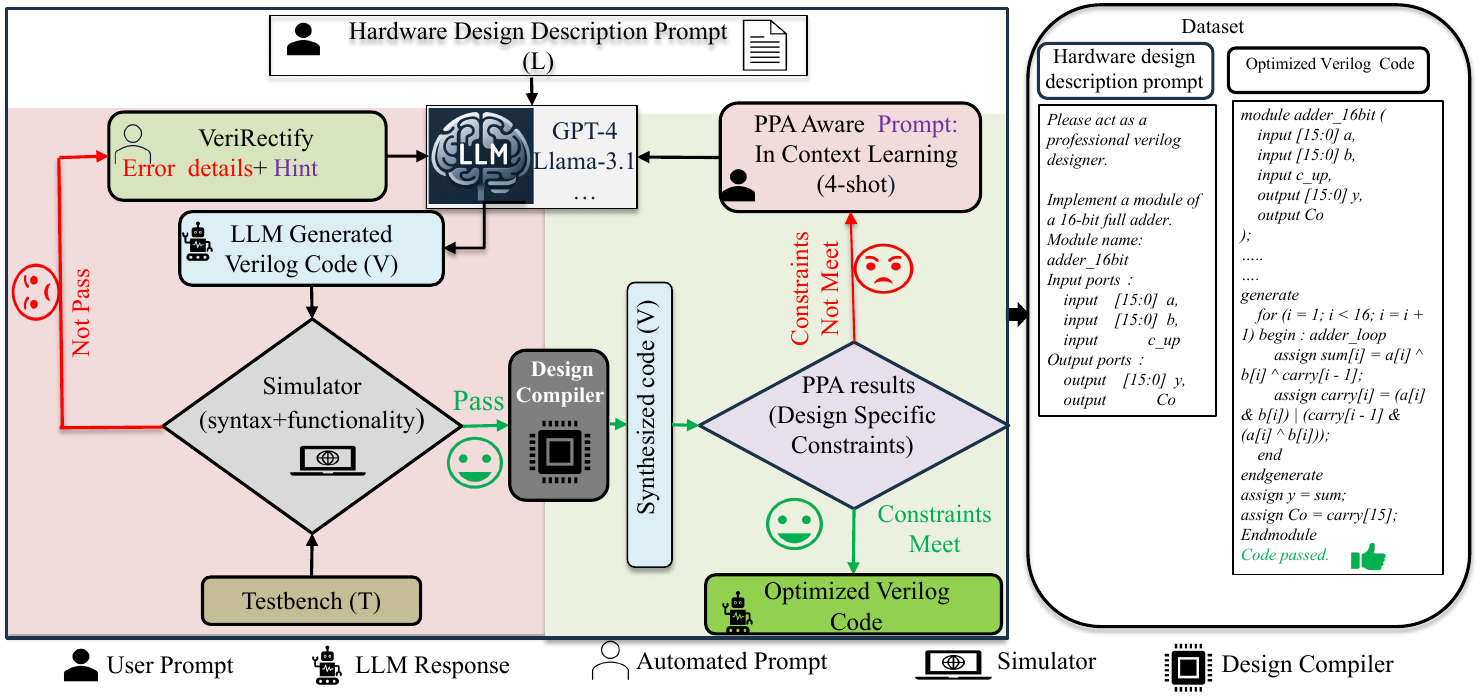} 
    \caption{VeriPPA framework.}
    \label{figure:Code_generation}
\end{figure*}

Despite various efforts, optimizing PPA remains the most critical task in chip design, and to the best of our knowledge, no existing methods support PPA optimization. Before we perform PPA optimization, we must generate correct Verilog code. Recent work in correct Verilog code generation falls into two categories: prompt engineering and fine-tuning. Prompt engineering improves Verilog code generation by adjusting descriptions and prompt structures. For example, hierarchical prompting \cite{10.1145/3670474.3685964} generates hierarchical code, ChipGPT \cite{chang2023chipgptfarnaturallanguage} applies prompt engineering for automatic chip generation, and RTLLM \cite{lu2023rtllm} uses self-planning prompt engineering to enhance correctness. Fine-tuning improves Verilog code generation by modifying model parameters. VeriGen \cite{thakur2023verigenlargelanguagemodel} uses fine-tuning on a collected dataset from GitHub, but lacks data cleaning and task-specific training, which reduces functional accuracy. ChipNeMo \cite{liu2024chipnemodomainadaptedllmschip} performs a two-round fine-tuning with in-house data, although only the first round benefits RTL code generation. BetterV \cite{pei2024bettervcontrolledveriloggeneration} fine-tunes the model alongside a generative discriminator, which increases deployment complexity. VerilogEval \cite{liu2023verilogeval} and RTLCoder \cite{10720939} provide benchmark datasets for single-round fine-tuning. 
In summary, no existing work addresses PPA optimization a very crucial aspect of chip design. There are two existing works that provide the PPA result (Vanilla) without optimization. In regards to the correct Verilog generation existing methods do not use the exact error details from the integrated simulator and multi-round conversation to understand the Verilog code for the circuit design. We highlighted the recent work in the comparison our \framework framework in Table \ref{tab:work_comparison}.
In this work, we propose \framework, a systematic open-source framework that makes LLM capable of PPA optimization and strengthens LLM's capability generation of Verilog code, as 
 shown in Figure~\ref{figure:Code_generation}.  
Our key contributions are summarized here:
\begin{itemize}
\item We introduce PPA optimizations to ensure that the generated Verilog codes meet design specification (i.e., PPA) is optimized, we use Synopsys Design Compiler to
perform logic synthesis and use the open source ASAP 7nm Predictive PDK to obtain PPA reports.
\item We propose an effective method for Verilog code generation using refinement of the errors by enabling the LLM to understand Verilog code for circuit design. We use the detailed error diagnostics from the iverilog simulator~\cite{Williams2023}, and pinpoint the exact location of syntactic or functional discrepancies as indicated by testbench failures as new prompts. We use multi-round generation to enhance the syntax and functionality correctness.
\item 
We incorporate in-context learning (ICL) in the  PPA domain to improve the LLM's understanding of PPA optimizations, especially when labeled data are scarce. By carefully creating diverse Verilog to PPA report pairs using different optimization strategies. Further, we create PPA aware prompt and corresponding strategy testbench for the Verilog design.

Compared with state-of-the-arts (SOTAs), e.g., RTLLM~\cite{lu2023rtllm}, VerilogEval~\cite{liu2023verilogeval}, our \framework achieves a success rate of  62.0\% (+16\%) for functional accuracy and 81.37\% (+8.3\%) for syntactic correctness in Verilog code generation on RTLLM dataset.
On the VerilogEval dataset,
our framework achieves 99.56\% syntactic correctness and 43.79\% functional correctness
surpassing current SOTA methods.


\begin{table}[ht]
    \centering
     \caption{Comparison of PPA and Verilog code generation using LLM works .}
    \label{tab:work_comparison}
    \resizebox{\columnwidth}{!}{ 
    \begin{tabular}{|c|c|c|c|}

    \hline
    \textbf{Work} & \textbf{Vanilla PPA} & \textbf{PPA Optimization} & \textbf{Accuracy} \\
    \hline
     \hline
    VeriGen \cite{thakur2023verigenlargelanguagemodel} & No & No & Moderate \\
    \hline
    RTLCoder \cite{10720939} & No & No & High\\
    \hline
   
    ChipNeMo \cite{liu2024chipnemodomainadaptedllmschip} & No & No & Moderate \\
    \hline 
   VerilogEval~\cite{liu2023verilogeval} & No & No & Moderate \\
    \hline
     BetterV \cite{pei2024bettervcontrolledveriloggeneration} & No & No & High \\
    \hline
    Revisiting VerilogEval\cite{pinckney2024revisiting} & No & No & High \\
    \hline
    
    RTLLM \cite{lu2023rtllm} & Yes & No & Moderate \\
    \hline
    ChipGPT \cite{chang2023chipgptfarnaturallanguage} & Yes & No & Moderate \\
    \hline
    VerilogCoder \cite{ho2025verilogcoderautonomousverilogcoding} & No & No & High \\
    \hline
   LLM-VeriPPA (Our Work) & \textbf{Yes} & \textbf{Yes} & \textbf{High} \\
    \hline
    \end{tabular}
    }

\end{table}

\end{itemize}

 \section{Backgound and Related Work}

Register Transfer Logic (RTL) is an critical abstraction in chip design that outlines how data moves between logical operations and registers. RTL is typically described using HDLs such as Verilog. In modern chip design workflows, human engineers manually convert design specifications into HDL before synthesizing them into circuits \cite{Blocklove_2023}. This manual translation process is time-consuming and susceptible to errors, which can lead to potential flaws in the hardware circuit designs. Recent advancements in Artificial Intelligence (AI), particularly LLMs, have enabled the automation of translating design specifications into HDL by understanding the instructions and generating codes in HDLs such as Verilog. The ability to generate HDL that meets specific design requirements, such as PPA, is crucial in chip design.

\noindent \textbf{Finetune LLMs.}
The LLMs have demonstrated various capabilities such as comprehension, reasoning, instruction following, and coding \cite{minaee2024largelanguagemodelssurvey}. However, their capability in generating practical hardware Verilog codes is limited because of insufficient available Verilog codes due to propriety natures of circuit designs \cite{fu2025gpt4aigchipnextgenerationaiaccelerator}. To address these challenges researches fine-tuned LLMs on hardware datasets.
Thakur \textit{et al.}~\cite{thakur2023benchmarking} advocate for the fine-tuning of open-source LLMs such as CodeGen~\cite{nijkamp2022codegen} to specifically generate Verilog code tailored for target designs. Subsequently, Chip-Chat~\cite{blocklove2023chip} delves into the intricacies of hardware design using LLMs,
highlighting the markedly superior performance of ChatGPT compared to other open-source LLMs.
Chip-GPT~\cite{chang2023chipgpt} also focuses on the task of register-transfer level (RTL) design by leveraging the capabilities of ChatGPT. 
However, these works mainly target the scale of simple and small circuits (e.g., \( < 20 \) designs with a average of \( < 45 \) Verilog code lines), as pointed out in~\cite{lu2023rtllm}.

\noindent \textbf{Enrich Verilog Source.} 
Several recent efforts focus on enriching Verilog codes.
RTLLM~\cite{lu2023rtllm}
introduces a benchmarking framework consisting of 30 designs that are specifically aimed at enhancing the scalability of benchmark designs. Furthermore, it utilizes effective prompt engineering techniques to improve the generation quality.
MG-Verilog \cite{zhang2024mgverilogmultigraineddatasetenhanced} provides the multi-level descriptions along side with code sample but its reliance on  Llama-2-70B-chat \cite{touvron2023llama} for annotation raises quality concerns about the dataset.
VerilogEval~\cite{liu2023verilogeval} assesses the performance of LLM in the realm of Verilog code generation for hardware design and verification. It comprises 156 problems from the Verilog instructional website HDLBits.  However, VerilogEval~\cite{liu2023verilogeval} does not offer PPA analysis for the generated codes. In RTLLM, the generated Verilog codes are directly extracted and synthesized using commercial tools to obtain PPA results, without PPA constraint-based feedback. 
Thus they suffer from limited generation quality.

\noindent \textbf{Verilog Code Agents}  VerilogCoder \cite{ho2025verilogcoderautonomousverilogcoding} introduces the multiple autonomous AI agents based on Abstract Syntax Tree (AST) based waveform tracing, graph planner, and other tools. These highly domain specific AI agent's output does not work on smaller LLMs and models goes to hallucination.  Further, It does not give provide any PPA optimizations.

ß

 \section{Framework}

\subsection{Design Overview}

In our  \framework framework, as illustrated in Figure~\ref{figure:Code_generation}, we use a text-based description ($.txt$ file) of hardware design, designated as $L$, 
to serve as input/prompt for the LLMs. 
$L$ 
details the module name, and specifies both input and output signals with the corresponding bit widths. 
In the first stage, highlighted in {light red}, we use LLM to parse the text-based description $L$ and generate the corresponding Verilog code $V$. $V$ is then subjected to syntax and functionality checks using the ICARUS Verilog simulator \cite{Williams2023} and a design-specific testbench $T$. Should $V$ fail these checks, we utilize VeriRectify (Section~\ref{sec:VeriRectify}) to provide an automated prompt to the LLM to correct the errors. If $V$ passes, it is synthesized to evaluate the Power, Performance, and Area (PPA) of the design.
The second stage, highlighted in {light green}, assesses the design-specific PPA requirements. We compare the PPA metrics of the synthesized code (after design compiler)
against the design constraints. If 
not meeting these constraints, a PPA-aware prompt (using in context learning) is fed back into the LLM for further optimization. Otherwise, it is saved as part of the dataset. 
The details of each technique are described in the subsequent subsections.



\subsection{Code Generation and Testing}
\vspace{-1pt}
\framework incorporates the ICARUS Verilog simulator \cite{Williams2023} to automate the evaluation (testing) of the generated codes. 
In contrast to high-level program languages such as Python, Verilog requires the use of testbenches, $T=\{T_1,T_2,…,T_m\}$, 
to systematically assess the code's functionality, encompassing a wide array of test scenarios.
Integrating the ICARUS Verilog simulator  into \framework provides immediate feedback on the code's syntactical and operational integrity.  The ICARUS Verilog simulator could pinpoint the exact location of syntactic errors or functional fails based on testbench test case failures. 
This integrated approach contrasts with frameworks such as RTLLM \cite{lu2023rtllm}, where an external simulator is used to check the correctness of the generated Verilog codes. 
\subsection{VeriRectify}
\label{sec:VeriRectify}

 \begin{figure}[ht]
    \centering
    \includegraphics[width=0.95\linewidth]{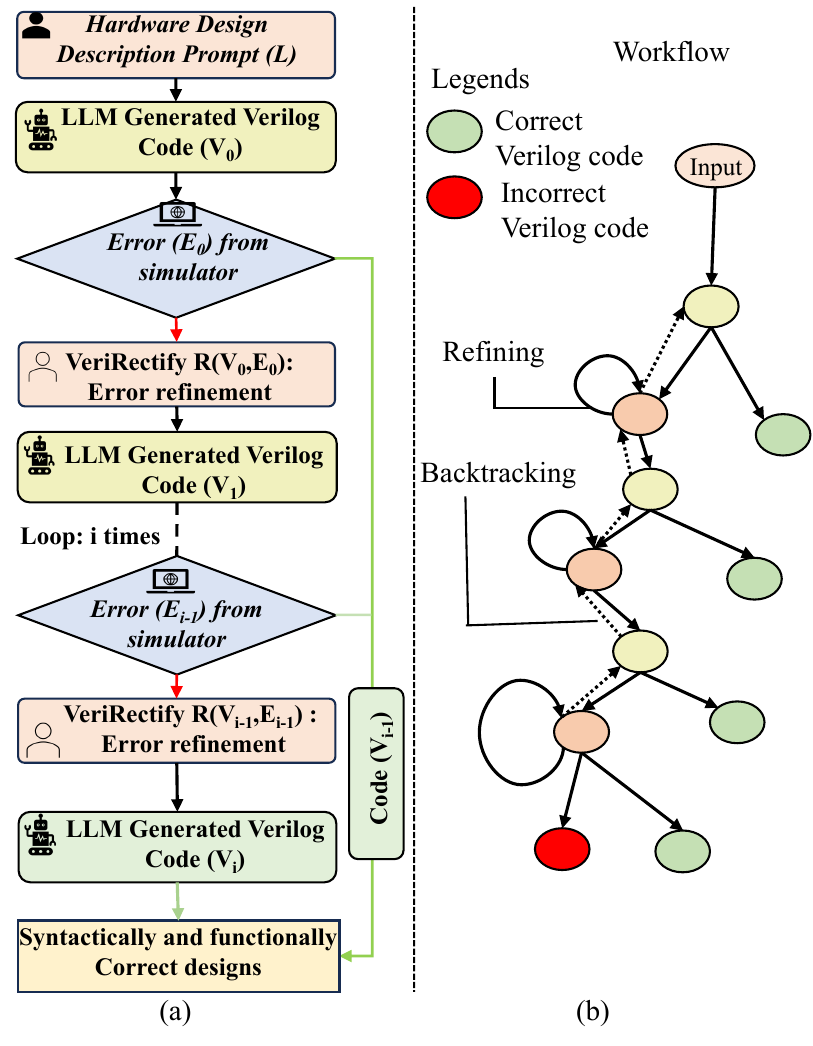} 
    \caption{(a) Multi-round conversation with error
feedback; (b) Workflow of the process.
    }
    \label{fig:multirectify}
\end{figure}
We
create a multi-iteration dialogue with an error feedback mechanism (Figure \ref{fig:multirectify} (a) and (b)), analogous to human problem-solving techniques. This method is designed as a recursive function that improves the output by carefully analyzing and correcting the errors found in previous iterations.  Let \(V_i\) denote the Verilog code resultant from the \(i^{\text{th}}\) iteration, and \(E_i\) represent the associated set of identified errors at this stage. Initially, \(V_0\) is the first generated code accompanied by its detected errors \(E_0\). Then the refinement function, \(R(V_i, E_i)\), which takes as input \(V_i\) and \(E_i\), and yields an enhanced code version \(V_{i+1}\) as output. Simultaneously, an error detection function \(D(V_i)\) is employed to identify errors within \(V_i\), generating 
\(E_i\).  The iterative process can be viewed as follows:\begin{equation} \label{eq:chain_of_thought}
  V_{i+1} = R(V_i, E_i) \quad \text{and} \quad E_{i+1} = D(V_{i+1})
\end{equation}



This process repeats until either no errors are detected or a predefined iteration limit, $K$ is reached, i.e., the iteration halts if, 
$D(V_{i+1})$=$\emptyset$ 
or 
$i=K$. 
$K$ is
empirically
adjustable (say 4)  based on observed results of code generation.
Thus, the multi-round conversation method  enhances code quality with each iteration until an optimal or satisfactory solution is reached within the bounds of 
$K$.
in this context is a systematic, iterative algorithm aimed at progressively minimizing the error in the generated Verilog code, enhancing code quality with each iteration until an optimal or satisfactory solution is reached within the bounds of 
$K$.



\subsection{Power Performance and Area (PPA).}
RTL simulation does not guarantee that the design (generated Verilog code) meet the design specification after we fabricate. Furthermore, the quality of the hardware design must be measured by its power, performance, and area metrics.

Our approach takes a step further by inspecting PPA of the design $V$ which passes the \emph{VeriRectify} process as the following:
\begin{equation} \label{eq:PPA}
\footnotesize
V = \begin{cases} 
V & \text{if } \text{PPA}(V) \text{ satisfies}, \\
\begin{aligned}
&\text{VeriRectify}
(V, \text{PPA}(V))
\end{aligned}
& \text{otherwise.}
\end{cases} 
\end{equation}

Our PPA check calls Synopsys Design Compiler to perform logic synthesis (and technology mapping) on the open-source ASAP 7nm Predictive PDK~\cite{vashishtha2017asap7}. We check all designs' warning/error messages during the logic synthesis, and the power ($\mu$W), area ($\mu m^2$), and clock (ps) for quality.
When the Verilog design can be synthesized and meets the PPA goal, it results in a pass. Otherwise, both the design and its corresponding PPA report will be fed back to the VeriRectify (Section~\ref{sec:VeriRectify}) for refinement.

The aim of PPA checking is to ensure the created design operates within a reasonable clock period, with acceptable power and area. This requires determining the power and area under optimal timing, or the smallest clock period.
\begin{figure*}[t]
    \centering
      \includegraphics[width=0.9\linewidth]{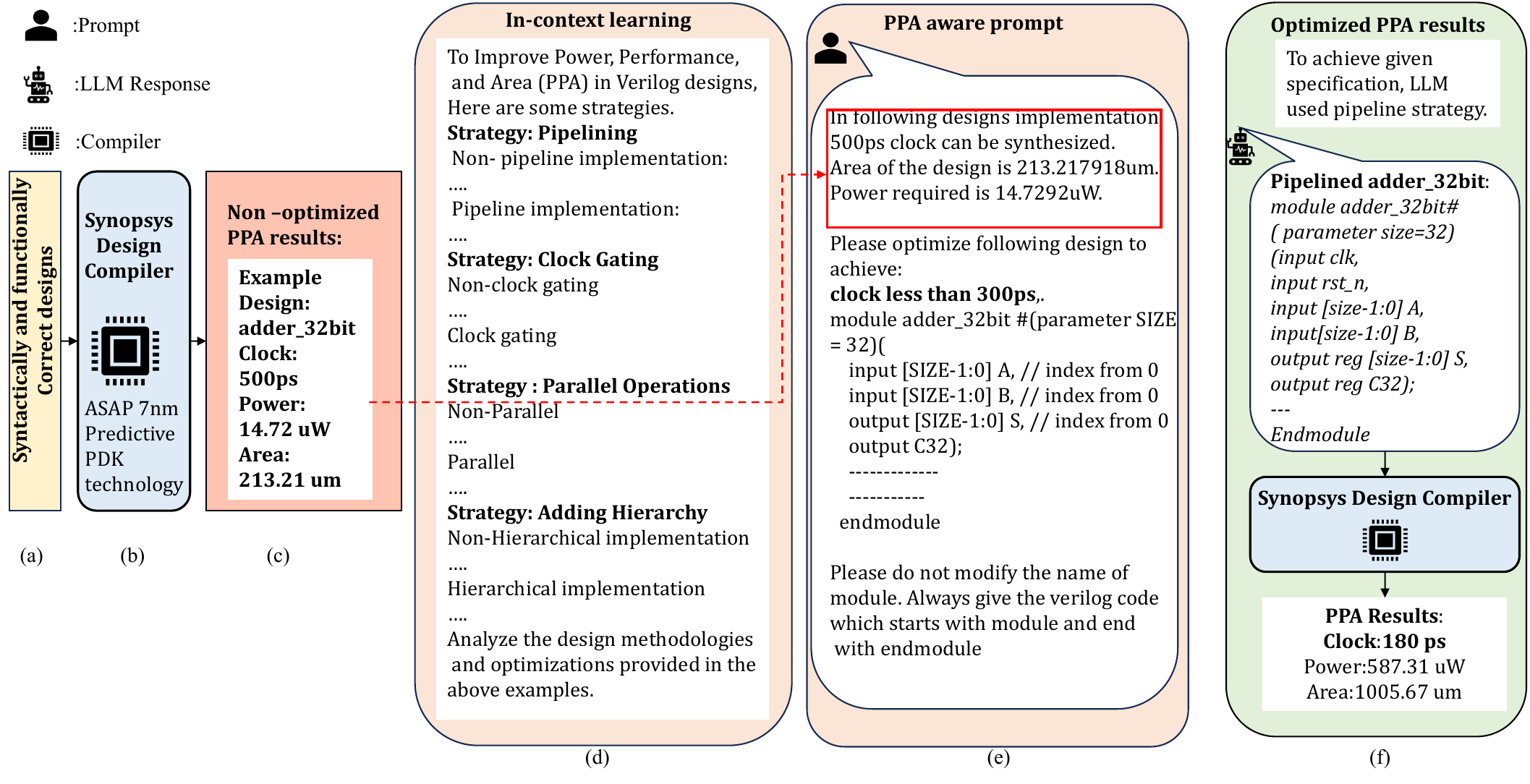}
      \vspace{-10pt}
  \caption{Optimization Flow; (a) Syntactically and functionally correct designs, (b) Synopsis compiler, (c) Non-optimized PPA results based on 7nm ASAP technology, (d) In-context learning to optimize PPA, (e) PPA aware prompt, (e) Optimized results}
    \label{fig:optimization_flow}
\end{figure*}

\section{Evaluation}

\subsection{Datasets}
In assessing our \framework framework, 
We utilize two benchmark datasets, the RTLLM dataset \cite{lu2023rtllm} includes 29 designs,
and the VerilogEval dataset \cite{liu2023verilogeval}, which comprises two subsets: VerilogEval-human, featuring 156 designs, and VerilogEval-machine, consisting of 108 designs.

\subsection{Experimental Setup}
We demonstrate the effectiveness of  \framework  for generating PPA-optimized Verilog for the given designs. We adopt GPT-3.5 \cite{openai_gpt3_5}, GPT-4 \cite{openai_gpt4}, GPT-4o \cite{openai2024gpt4o}, Llama-2-7B \cite{touvron2023llama}, Llama-3-8B \cite{meta_llama3}, Codellama-7B \cite{roziere2023code}, Llama3.1-405B \cite{MetaLLama3.1}, RTL-Coder \cite{liu2024rtlcoderfullyopensourceefficient}, and DeepSeek Coder \cite{guo2024deepseekcoderlargelanguagemodel} as our LLM models. We use $n$=1, temperature temp = 0.7, and a context length of 2,048. Further, we incorporate the ICARUS Verilog simulator \cite{Williams2023} to automate the testing of the generated code. 
For PPA check, we perform the logic synthesis using Synopsys Design Compiler with \texttt{compile\_ultra} command and we use the ASAP 7nm Predictive PDK~\cite{vashishtha2017asap7}. 
We implement an in-house simulator to sweep the timing constraints to find the fastest achievable clock frequency for all the generated designs. 
All experiments are conducted on a Linux-
based host with AMD EPYC 7543 32-Core Processor and an
NVIDIA A100-SXM 80 GB.

\subsection{Generation Correctness}
We evaluate Verilog generation accuracy using two primary metrics: syntax checking and functionality verification. Table~\ref{gpt} shows results from our methodology of correcting Verilog code through multiple correction attempts. For each design description, five codes are generated, with up to four corrections per generation. The number of correction attempts is set to four because after this point, correction efficiency decreases due to repetitive model outputs.
For the GPT-3.5 model, initial syntax correctness is 44.13\% and functionality correctness is 24.13\%. Applying the \framework framework changes syntax correctness to 65.51\% and functionality correctness to 31.03\%. For the RTLLM baseline, syntax correctness is 32.41\% and functionality correctness is 20.68\%. With the \framework framework, these become 55.17\% and 31.03\%.
For the GPT-4 model, syntax correctness is 66.20\% and functionality correctness is 37.93\%. With \framework, syntax correctness is 81.37\% and functionality correctness is 48.27\%. RTLLM baseline scores are 60.00\% for syntax and 34.48\% for functionality. After applying \framework, these become 77.93\% and 48.27\%.
Testing GPT-4 with four-shot learning, syntax correctness is 70.34\% and functionality correctness is 37.93\%. With \framework, syntax correctness is 79.31\% and functionality correctness is 41.37\%. For RTLLM, syntax correctness is 66.89\% and functionality correctness is 44.82\%. With \framework, syntax correctness is 81.37\% and functionality correctness is 62.06\%.
For GPT-4o, syntax correctness is 75.17\% and functionality correctness is 44.82\%. With \framework, syntax correctness is 86.20\% and functionality correctness is 48.27\%. RTLLM baseline scores are 75.86\% for syntax and 41.37\% for functionality. After applying \framework, syntax correctness is 82.06\% and functionality correctness is 44.82\%.

These results show that applying the \framework framework to both GPT and RTLLM models changes both syntax and functionality correctness across all tested models and settings.

We evaluated VeriPPA using open-source LLMs, demonstrating its strong effectiveness with these models. Starting with the large Llama 3.1-405B model \cite{MetaLLama3.1}, VeriPPA increases syntax accuracy from 31.72\% to 80.68\% and functionality accuracy from 20.68\% to 44.82\%, as shown in Table \ref{llama}. For smaller models, VeriPPA continues to enhance syntax correctness. With Llama 2-7B \cite{touvron2023llama} , syntax correctness rises from 20.68\% to 27.58\%. In the case of Llama 3-8B \cite{meta_llama3}, syntax correctness improves from 3.4\% to 17.5\%. Codellama-7B \cite{roziere2023code} shows an increase in syntax correctness from 16.2\% to 28.35\%. Despite these syntax improvements, these smaller models do not achieve functionality correctness due to the rigorous tests in our test benches. As demonstrated in Table \ref{llama}, VeriPPA is highly effective with larger open-source LLMs.

\begin{table*}[t]
\begin{center}
\caption{\label{gpt} Comparison of Different Models and RTLLMs methods on RTLLM dataset.}
\resizebox{0.99\linewidth}{!}{
\begin{tabular}{|c||c|c|c|c||c|c|c|c|}
\hline
\multirow{2}{*}{\textbf{Model}} & \multicolumn{2}{c|}{\textbf{Syntax (\%)}} & \multicolumn{2}{c||}{\textbf{Functionality (\%)}} & \multicolumn{2}{c|}{\textbf{Syntax (\%)}} & \multicolumn{2}{c|}{\textbf{Functionality (\%)}} \\
\cline{2-9}
 & \textbf{w/o VeriPPA} & \textbf{w/ VeriPPA} & \textbf{w/o VeriPPA} & \textbf{w/ VeriPPA} & \textbf{RTLLM } & \textbf{RTLLM w/ VeriPPA} & \textbf{RTLLM } & \textbf{RTLLM w/ VeriPPA} \\
\hline \hline
GPT-3.5 & 44.13 & 65.51 & 24.13 & 31.03 & 32.41 & 55.17 & 20.68 & 31.03 \\
\hline
GPT-4 & 66.20 & 81.37 & 37.93 & 48.27 & 60.00 & 77.93 & 34.48 & 48.27 \\
\hline
GPT-4 (4-shot) & 70.34 & 79.31 & 37.93 & 41.37 & 66.89 & 81.37 & 44.82 & 62.06 \\
\hline
GPT-4o & 75.17 & 86.20 & 44.82 & 48.27 & 75.86 & 82.06 & 41.37 & 44.82 \\
\hline
\end{tabular}}
\end{center}
\vspace{-15pt}
\end{table*}


\begin{table}[t]
\begin{center}
\caption{\label{llama} Open source LLM Results.}
\resizebox{0.98\linewidth}{!}{
\begin{tabular}{|c||c|c|c|c|} 
 \hline 
 \textbf{Model} & \multicolumn{2}{c|}{\textbf{Without VeriPPA}} & \multicolumn{2}{c|}{\textbf{With VeriPPA}} \\
 \hline 
  & \textbf{Synt. (\%)} & \textbf{Funct. (\%)} & \textbf{Synt. (\%)} & \textbf{Funct. (\%)} \\
 \hline \hline
Llama3.1-405B & 31.72  & 20.68  & 80.68  &44.82 \\
\hline
Llama-2-7B & 20.68 & 0 & 27.58 & 0 \\
 \hline
 Llama-3-8B& 3.4 & 0 & 17.24 & 0 \\
 \hline
 CodeLlama  7B & 16.2  & 0  & 28.35 & 0 \\
  \hline
\end{tabular}}
\end{center}
\vspace{-18pt}
\end{table}

We evaluate the \framework framework using the VerilogEval-Machine and VerilogEval-Human datasets. Table~\ref{verilogeval} summarizes results for VerilogEval-Machine. For GPT-4, syntax correctness with Revisiting VerilogEval~\cite{pinckney2025revisitingverilogevalyearimprovements} is 92.11\%, and with VeriPPA it is 99.56\%. Functionality correctness for GPT-4 is 33.57\% for Revisiting VerilogEval and 43.79\% for VeriPPA. Using GPT-4 with four-shot learning, syntax correctness is 90.21\% (Revisiting VerilogEval) and 95.91\% (VeriPPA), while functionality correctness is 35.76\% and 45.25\%, respectively. For the VerilogEval-Human dataset, as shown in Table~\ref{verilogevalhuman}, GPT-4 syntax correctness is 91.28\% with Revisiting VerilogEval and 97.17\% with VeriPPA. Functionality correctness is 29.48\% (Revisiting VerilogEval) and 39.74\% (VeriPPA). The four-shot learning variant of GPT-4 shows similar results, with syntax correctness at 88.97\% and 95.76\%, and functionality correctness at 29.4\% and 39.74\% for Revisiting VerilogEval and VeriPPA, respectively.
We also evaluate the \framework framework using RTL-Coder~\cite{liu2024rtlcoderfullyopensourceefficient} and DeepSeek Coder~\cite{guo2024deepseekcoderlargelanguagemodel}. For DeepSeek Coder on the VerilogEval-Machine dataset, syntax correctness is 55.12\% with Revisiting VerilogEval and 78.97\% with VeriPPA. Functionality correctness is 16.66\% (Revisiting VerilogEval) and 24.35\% (VeriPPA). For RTL-Coder, syntax correctness is 0.38\% with Revisiting VerilogEval and 27.94\% with VeriPPA, while functionality correctness is 0.64\% and 1.28\%, respectively. These results show that applying VeriPPA with the \framework framework changes both syntax and functionality correctness across different models and datasets. Using four-shot learning also changes functionality correctness, indicating the benefit of multi-sample correction for Verilog code generation.

\begin{table}[t]
\begin{center}
\caption{\label{verilogeval} Comparison on VerilogEval-Machine dataset: Revisiting VerilogEval~\cite{pinckney2025revisitingverilogevalyearimprovements} vs. VeriPPA.}
\resizebox{0.98\linewidth}{!}{
\begin{tabular}{|c||c|c||c|c|}
\hline
\multirow{2}{*}{\textbf{Model}} & \multicolumn{2}{c||}{\textbf{Revisiting VerilogEval~\cite{pinckney2025revisitingverilogevalyearimprovements}}} & \multicolumn{2}{c|}{\textbf{VeriPPA}} \\
\cline{2-5}
 & \textbf{Syntax (\%)} & \textbf{Function (\%)} & \textbf{Syntax (\%)} & \textbf{Function (\%)} \\
\hline \hline
GPT-4 & 92.11 & 33.57 & 99.56 & 43.79 \\
\hline
GPT-4 (4-shot) & 90.21 & 35.76 & 95.91 & 45.25 \\
\hline
RTL-Coder & 0.38 & 0.64 & 27.94 & 1.28 \\
\hline
DeepSeek-Coder-67B & 55.12 & 16.66 & 78.97 & 24.35 \\
\hline
\end{tabular}}
\end{center}
\vspace{-15pt}
\end{table}

\begin{table}[t]
\begin{center}
\caption{\label{verilogevalhuman} Comparison on VerilogEval-Human dataset: Revisiting VerilogEval~\cite{pinckney2025revisitingverilogevalyearimprovements} vs. VeriPPA.}
\resizebox{0.98\linewidth}{!}{
\begin{tabular}{|c||c|c||c|c|}
\hline
\multirow{2}{*}{\textbf{Model}} & \multicolumn{2}{c||}{\textbf{Revisiting VerilogEval~\cite{pinckney2025revisitingverilogevalyearimprovements}}} & \multicolumn{2}{c|}{\textbf{VeriPPA}} \\
\cline{2-5}
 & \textbf{Syntax (\%)} & \textbf{Function (\%)} & \textbf{Syntax (\%)} & \textbf{Function (\%)} \\
\hline \hline
GPT-4 & 91.28 & 29.48 & 97.17 & 39.74 \\
\hline
GPT-4 (4-shot) & 88.97 & 29.48 & 95.76 & 39.74 \\
\hline
\end{tabular}}
\end{center}
\end{table}
\subsection{PPA Optimization}
\begin{table}[t]
\begin{center}
\caption{\label{PPA_table} PPA results of generated Verilog code }
\resizebox{0.48\textwidth}{!}{
\begin{tabular}{|c| c | c | c | c | c |c|} 
 \hline 
 \textbf {Design Name} & \multicolumn{3}{c||}{\textbf{GPT-4}} & \multicolumn{3}{c||}{\textbf{GPT-4 (4-shot)}}\\ 
 \hline 
  & \multicolumn{1}{c|}{\begin{tabular}{@{}c@{}}Clock \\ (ps)\end{tabular}} & \multicolumn{1}{c|}{\begin{tabular}{@{}c@{}}Power \\ (\(\scriptstyle\mu\)W)\end{tabular}} & \multicolumn{1}{c||}{\begin{tabular}{@{}c@{}}Area \\ (\(\scriptstyle\mu\)m\(^2\))\end{tabular}} & \multicolumn{1}{c|}{\begin{tabular}{@{}c@{}}Clock \\ (ps)\end{tabular}} & \multicolumn{1}{c|}{\begin{tabular}{@{}c@{}}Power \\ (\(\scriptstyle\mu\)W)\end{tabular}} & \multicolumn{1}{c||}{\begin{tabular}{@{}c@{}}Area \\ (\(\scriptstyle\mu\)m\(^2\))\end{tabular}}  \\
 \hline \hline
 adder\_8bit & 318.5 & 6.3 & 38.5 & 333.1 & 6.1 & 42.9 \\
 \hline
 adder\_16bit & 342.2 & 10.9 & 104.5 & 135.1 & 41.1 & 152.8 \\
 \hline
 adder\_32bit & 500.0 & 14.2 & 211.6 & 500.0 & 14.7 & 213.2 \\
 \hline
 multi\_booth & 409.0 & 112.1 & 526.0 & 409.0 & 112.1 & 526.0 \\
 \hline
 right\_shifter & 47.5 & 144.3 & 42.9 & 47.5 & 144.3 & 42.9 \\
 \hline
 width\_8to16 & 74.1 & 223.2 & 145.8 & 145.6 & 128.7 & 157.2 \\
 \hline
 edge\_detect & 61.5 & 49.0 & 23.3 & 61.5 & 49.0 & 23.3 \\
 \hline
 mux & 54.7 & 215.3 & 86.1 & 54.7 & 215.3 & 86.1 \\
 \hline
 pe & 500.0 & 552.5 & 2546.5 & 500.0 & 541.0 & 2488.6 \\
 \hline
 asyn\_fifo & 295.2 & 406.4 & 1279.3 & 228.3 & 526.6 & 1295.4 \\
 \hline
 counter\_12 & 134.4 & 33.1 & 40.6 & 124.5 & 34.6 & 36.4 \\
 \hline
 fsm & 88.3 & 32.7 & 31.5 & 68.7 & 49.0 & 50.2 \\
 \hline
 multi\_pipe\_4bit & 254.7 & 40.7 & 131.3 & - & - & - \\
 \hline
 pulse\_detect & 10.3 & 187.5 & 13.5 & 32.7 & 59.1 & 13.5 \\
 \hline
 calendar & - & - & - & 208.6 & 86.6 & 199.0 \\
 \hline
\end{tabular}}
\end{center}
\vspace{-10pt}
\end{table}

In this section, we shift focus from verifying the correctness of the generated Verilog codes to optimizing its quality.
In  \framework, 
We use the Synopsys Design Compiler to synthesize our designs and generate PPA reports. 
Table \ref{PPA_table} shows the results of different designs with different LLM models. 
It shows the without optimization PPA (Vanilla) result of each design. To demonstrate \framework ability to optimize PPA later, we show best Vanilla PPA from multiple PPA reports.
For example, the $pulse\_detect$ design passes five times functionally and syntactically. Therefore, in post-synthesis, we collect five PPA reports for the $pulse\_detect$ design, and we select the best PPA result to include in Table \ref{PPA_table}. However, these best PPA results do not meet design-specific PPA requirements,

 
To address this, 
We further perform the PPA constraint-based feedback mechanism, integrated with ICL, as illustrated in Figure \ref{fig:optimization_flow}. 
This approach represents a significant step towards aligning LLM-generated codes with application-specific PPA requirements.
 {Figure \ref{fig:optimization_flow} demonstrates our process, starting with the collection of syntactically and functionally correct designs and generating non-optimized PPA results as shown in Figure \ref{fig:optimization_flow} (a), (b), and (c).
The non-optimized PPA results do not meet application-specific PPA requirements. For example, $adder\_32bit$, can be synthesized with a 500ps clock as shown in Figure \ref{fig:optimization_flow} (c). However, this clock speed does not fulfill the rapid clock requirements necessary for some applications, such as cryptographic hardware which consists of adders, where a fast clock is crucial, but area and power constraints are less critical.
To enhance the speed of $adder\_32bit$, we impose a clock constraint, aiming for a clock speed of less than 300ps, as outlined in the PPA constraint-based prompt in Figure \ref{fig:optimization_flow} (e). The framework instructs the LLM to consider various optimization strategies, including Pipelining, Clock Gating, Parallel Operation, and Hierarchical Design as depicted in Figure \ref{fig:optimization_flow} (d).
Upon providing the PPA-based constraint prompt and context to the LLM, we analyze the resultant Verilog code for syntax and functional accuracy, making corrections where necessary. If the code passes both checks, we proceed to its final synthesis, achieving an optimized Verilog code as shown in Figure \ref{fig:optimization_flow} (f), where the $adder\_32bit$ operates at an improved 180ps clock. In Table \ref{PPA_table_optimized}, we present the results of selected optimized designs. 
Notably, no design from the VerilogEval \cite{liu2023verilogeval} dataset is present in Table \ref{PPA_table_optimized}, as those designs did not require complex optimization.

\begin{table}[t]
\begin{center}
\caption{\label{PPA_table_optimized} PPA Optimized Verilog Design Results}
\resizebox{0.9\linewidth}{!}{
\begin{tabular}{|c||c|c|c|} 
 \hline 
 \textbf{Design Name} & \textbf{Clock (ps)} & \textbf{Power (\(\mu\)W)} & \textbf{Area (\(\mu\)m)}   \\
 \hline \hline
 adder\_32bit & 180.0 & 587.31 & 1005.67 \\
 \hline
 multi\_booth  & 123.2 & 42.39 & 42.92 \\
 \hline
 pe  & 325.0 & 1206.0 & 4863.88 \\
 \hline
 asyn\_fifo &  114.8 & 988.92 & 1344.86 \\
 \hline
\end{tabular}}
\end{center}
\end{table}
\subsection{Computational cost analysis between self-planning (SOTA) and VeriPPA}
To provide a fair comparison, we first limit both methods to a single iteration: the self-planning (SOTA) and our VeriPPA approach. 
As shown in Table \ref{table:A}, VeriPPA used 46,702 fewer tokens (a 14.71\% reduction) compared to SOTA, while also achieving better syntax accuracy 80.68\% for VeriPPA versus 77.93\% for SOTA, with the same functionality accuracy 41.37\%.  Our estimation shows, for a GPT-4o VeriPPA requires 95272.08 trillion less Multiply-Accumulate Operations (MACs) than the SOTA . 
Overall, results demonstrate that VeriPPA has lower computational costs than the SOTA method while maintaining or improving accuracy. 



\begin{table}[t]
\begin{center}
\caption{\label{table:A} Analysis Table - One iteration comparison using GPT-4o \cite{openai2024gpt4o} model}
\vspace{-4pt}
\resizebox{0.98\linewidth}{!}{
\begin{tabular}{|c||c|c|c|c|} 
 \hline 
 \multirow{2}{*}{\textbf{Method}} & \multirow{2}{*}{\textbf{MACs}} & \multirow{2}{*}{\textbf{Tokens}} & \multicolumn{2}{c|}{\textbf{Accuracy (\%)}} \\
 \cline{4-5}
  & & & \textbf{Syntax} & \textbf{Functionality} \\
 \hline \hline
Self-planning  & 647589.84 & 317446 & 77.93 & 41.37 \\
\hline
VeriPPA  & 552317.76 & 270744 & 80.68 & 41.37 \\
 \hline
\end{tabular}}
\end{center}
\end{table}

\section{Conclusion}

In this paper, we introduce a novel framework \framework, designed to assess and enhance LLM efficiency in this specialized area. Our method includes generating initial Verilog code using LLMs, followed by a unique two-stage refinement process. The first stage focuses on improving the functional and syntactic integrity of the code, while the second stage aims to optimize the code in line with Power-Performance-Area (PPA) constraints, an essential aspect of effective hardware design.
This dual-phase approach of error correction and PPA optimization has led to notable improvements in the quality of LLM-generated Verilog code. Our framework achieves 62.0\% (+16\%) for functional accuracy and 81.37\% (+8.3\%) for syntactic correctness in Verilog code generation, compared to SOTAs.

\bibliographystyle{plain}
\bibliography{bib/ref}
\end{document}